\documentclass[twocolumn]{article}

\usepackage{epsfig}
\usepackage{amssymb}

\newtheorem{corollary}{Corollary}
\newtheorem{theorem}{Theorem} 
\newtheorem{lemma}{Lemma}

\def\Proof{\par\noindent{\bf Proof:}\indent}
\def\QED{\hfill$\Box$\par\vskip1em}    

\begin{document}
\title{\emph{2FACE}: Bi-Directional Face Traversal for \\ 
       Efficient Geometric Routing}
\author{Mark Miyashita
\quad Mikhail Nesterenko\thanks{This author 
was supported in part by DARPA contract OSU-RF \#F33615-01-C-1901 and 
by NSF CAREER Award 0347485.}\\
Department of Computer Science \\
Kent State University\\
Kent, OH, 44240, USA\\
\{mmiyashi, mikhail\}@cs.kent.edu\\
 \\
Technical report TR-KSU-CS-2006-06, Kent State University\\
}

\maketitle

\begin{abstract}
We propose bi-directional face traversal algorithm $2FACE$ to shorten
the path the message takes to reach the destination in geometric
routing.  Our algorithm combines the practicality of the best
single-direction traversal algorithms with the worst case message
complexity of $O(|E|)$, where $E$ is the number of network edges.  We
apply $2FACE$ to a variety of geometric routing algorithms. Our
simulation results indicate that bi-directional face traversal
decreases the latency of message delivery two to three times compared
to single direction face traversal. The thus selected path approaches
the shortest possible route. This gain in speed comes with a similar
message overhead increase. We describe an algorithm which compensates
for this message overhead by recording the preferable face traversal
direction. Thus, if a source has several messages to send to the
destination, the subsequent messages follow the shortest route. Our
simulation results show that with most geometric routing algorithms
the message overhead of finding the short route by bi-directional face
traversal is compensated within two to four repeat messages.
\end{abstract}

\section{Introduction}
Geometric routing is an elegant approach to data dissemination in
resource-constrained and large-scale ad hoc networks such as wireless
sensor networks. Geometric routing is attractive because it does not
require nodes to maintain, or the messages to carry, extensive state
or routing information. 

In geometric routing, each node knows its own and its neighbors'
coordinates.  Using low-cost GPS receivers or location estimation
algorithms \cite{GPS}, wireless sensor nodes can learn their relative
location with respect to the other nodes and then use this information
to make routing decisions.  The message source node knows the
coordinates of the destination node. The information that the message
can carry should not depend on the network size. Each forwarding node
should not maintain any extended routing data or keep any information
about forwarded messages between message transmissions. The lack of
infrastructure makes geometric routing algorithm a popular
initialization and fallback option for other routing schemes.  Thus,
geometric routing optimization is of interest to the broad community
of wireless sensor network designers.

\ \\ \emph{Greedy routing} \cite{Finn} is an elementary approach to
geometric routing where the node selects the neighbor closest to the
destination and forwards the message there. The process repeats until
the destination is reached. Greedy routing fails in case the node is a
\emph{local minimum}: it does not have neighbors that are closer to
the destination than itself. Alternatively, in \emph{compass routing}
\cite{Compass}, a node selects the neighbor whose direction has the
smallest angle to the direction of the destination. Compass routing is
prone to livelocks.

GFG \cite{GFG}(also known as GPSR \cite{GPSR}) guarantees message
delivery. GFG contains two parts. A node uses greedy routing to
forward the message. To get out of \emph{local minimum}, GFG switches
to \emph{face traversal}. In GFG, the message sequentially traverses
faces that intersect the line from the source to destination node.
The face traversal always proceeds in a fixed direction clockwise or
counterclockwise. GFG may use two face traversal algorithms:
\emph{FACE-1} and \emph{FACE-2}. In \emph{FACE-1}, the message goes
around the entire face to find the adjacent face that is the closest
to the destination.  The message is then sent directly to this next
face and the process repeats. In \emph{FACE-2}, the message switches
faces as soon as it finds the next face to traverse. The new face may
not be closest to the destination and the message may have to
traverse the same face multiple times.  \emph{FACE-1} and
\emph{FACE-2} have the respective worst case message complexity of
$3|E|$ and $|V|^2$, where $V$ and $E$ are the set of vertices and
edges in the network graph. Even though \emph{FACE-2} has worse
message complexity, it proves to be more efficient in practice.

Datta et al \cite{Shortcut} propose a number of optimizations to GFG.
In particular, they propose that nodes maintain distance-two
non-planar neighbors. If these nodes lie on the same face, the edge
from the non-planar graph may be used as a shortcut to traverse the
face.  Kuhn et al \cite{TheoryPractice} modify GFG to achieve
asymptotically optimal worst-case message complexity.  Nesterenko and
Vora \cite{Void} propose a technique of traversing voids in non-planar
graphs similar to face traversal. This traversal may be combined with
greedy routing in $GVG$ similar to $GFG$. A number of geometric
(location-based) routing algorithms are proposed. The reader is
referred to the following survey for a comprehensive list
\cite{Taxonomy}.

One of the shortcomings of face and void traversal is the possibility
of producing a route that is far longer than optimal.  The problem
lies in the fixed traversal direction used in the existing
algorithms. When routing along a face, the route in one direction may
be significantly shorter than in the other. This is often the case
when the message has to traverse the external face of the graph.  We
propose an algorithm \emph{2FACE} that accelerates the message
propagation by sending the message to traverse the face in both
directions concurrently. When one of the message encounters a face
that is closer to the destination, the message spawns two messages to
traverse the new face and continues to traverse the old face. When the
two messages traveling around the face in the opposite direction meet,
the traversal stops. The node memory and message-size requirements of
\emph{2FACE} are the same as the other geometric routing algorithms.
\emph{2FACE} improves worst-case time and message complexity of
comparable single-direction algorithms.  In practice, \emph{2FACE}
guarantees faster message delivery to the destination but may require
more messages.  We present a technique to use \emph{2FACE} to
determine the intermediate nodes to learn the preferred traversal
direction. If source has multiple messages to send to the same
destination, this technique can be used to eliminate the message
overhead as the subsequent messages use the shorter path.

\ \\ \textbf{Paper contribution and organization.}  The rest of the
paper is organized as follows. We introduce our notation in
Section~\ref{secPrelim}. We describe \emph{2FACE} and formally prove
it correct in Section~\ref{secAlg}. In Section~\ref{secExt}, we
discuss how the algorithm can be adopted for greedy routing for use in
non-planar graphs and how algorithm can be used to select a preferred
route in multi-message sessions. We evaluate the performance of our
algorithm and its modifications in Section~\ref{secSim} and conclude
the paper in Section~\ref{secEnd}.

\section{Preliminaries} \label{secPrelim}
\textbf{Graphs.} We model the network as a connected geometric graph
$G=(V,E)$.  The set of \emph{nodes} (\emph{vertices}) $V$ are embedded
in a Euclidean plane and are connected by \emph{edges} $E$. The graph
is \emph{planar} if its edges intersect only at vertices. A
\emph{void} is a region on the plane such that any two points in this
region can be connected by a curve that does not intersect any of the
edges in the graph. Every finite graph has one infinite
\emph{external} void. The other voids are internal.  A void of a
planar graph is a \emph{face}.

\ \\ \textbf{Face traversal.} \emph{Right-hand-rule} face traversal
proceeds as follows. If a message arrives to node $a$ from its
neighbor $b$, $a$ examines its neighborhood to find the node $c$ whose
edge $(a,c)$ is the next edge after $(a,b)$ in a clockwise
manner. Node $a$ forwards the message to $c$. This mechanism results
in the message traversing an internal face in the counter-clockwise
direction, or traversing the external face in the clockwise
direction. \emph{Left-hand-rule} traversal is similar, except the
next-hop neighbor is searched in the opposite direction. If node $n$
(i) borders two faces $F$ and $F'$ both of which intersect the $(s,d)$
line and (ii) there is an edge adjacent to $n$ that borders $F$ and
$F'$ and intersects $(s,d)$, then $n$ is an \emph{entry point} to $F'$
and $F$. A source node is an entry point to the first face that
intersects $(s,d)$. Notice that according to this definition, both
nodes adjacent to the edge that intersects $(s,d)$ are entry
points. To simplify the presentation we assume that only one of them
is an entry point while the other is a regular border node. However,
\emph{2FACE} is correct without this assumption.  Notice that a face
may intersect $(s,d)$ in multiple places and thus have multiple entry
points. Two faces that share an entry point are \emph{adjacent}.

\ \\ \textbf{Geometric routing.} A source node $s$ has a message to
transmit to a destination node $d$. Node $s$ is aware of the Euclidean
coordinates of $d$. Node $s$ attaches its own coordinates as well as
those of $d$ to the messages. Thus, every node receiving the message
learns about the line $(s,d)$ that connects the source and the
destination. Each message is a \emph{token}, as its payload is
irrelevant to its routing. Depending on whether the token is routed
using right- or left-hand-rule, it is denoted as $R$ or $L$.  Each
node $n$ knows the coordinates of its \emph{neighbors}: the nodes
adjacent to $n$ in $G$.

\ \\ \textbf{Execution model.} We assume that each node can send only
one message at a time. The node does not have control as to when the
sent message is actually transmitted. After the node appends the
message to the send queue $SQ$, the message may be sent at arbitrary
time.  Each channel has zero capacity; that is, the sent message is
removed from $SQ$ of the sender and instantaneously appears at the
receiver. Message transmission is reliable. The node may examine and
modify $SQ$. We assume that $SQ$ manipulation, including its
modification and message transmission, is done \emph{atomically}. We
assume that the execution of the algorithm is a sequence of atomic
actions.  The system is \emph{asynchronous} in the sense that the
difference between algorithm execution speed at each process is
arbitrary (but finite).

\ \\ \textbf{Complexity measures.} The worst case message complexity
of an algorithm is the largest number of messages that is sent in a
single computation calculated in terms of the network parameters. The
worst case time complexity is the longest chain of causally related
messages in a computation. Where two messages are causally related if
the send of one message causally follows the receipt of the other.

\section{\emph{2FACE} Description and \\ Correctness Proof} \label{secAlg}

\textbf{Description.}  The pseudocode of \emph{2FACE} is shown in
Figure \ref{Fig2FACEcode}.  The operation of \emph{2FACE} is as
follows. The source $s$ initiates the face traversal by sending the
right- and left-hand rule tokens $R$ and $L$ to traverse the face $F$
that intersects the $(s,d)$ line. When a node $n$ receives token $L$
it first checks if it already has a matching $L$. If there is a
matching token, both tokens are removed and the processing stops. If
the node is the destination, the token is delivered. Note that for the
same of uniformity, the face traversal continues after delivery. If
$n$ is an entry point to the adjacent face $F'$, $n$ initiates the
traversal $F'$ by sending $L$ and $R$ tokens to go around $F'$. After
that, $n$ retransmits $L$.  Processing of the receipt of $R$ is
similar to that of $L$.

\begin{figure}
\centering
\begin{tabbing}
1234\=1234\=1234\=1234\=12345\=12345\=12345\=12345\=12345\=12345\=\kill

\textbf{node} $s$\\
\>/* let $F$ be the face bordering $s$ \\
\>\>and intersecting line $(s,d)$\\
\>\textbf{add} $L(s,d,F)$ to $SQ$\\
\>\textbf{add} $R(s,d,F)$ to $SQ$\\
\\
\textbf{node} $n$\\
\>\textbf{if} \textbf{receive} $L(s,d,F)$ \textbf{then}\\
\>\>\textbf{if} $R(s,d,F) \in SQ$ \textbf{then} \\
\>\>\>\textbf{delete} $L(s,d,F)$ from $SQ$\\
\>\>\textbf{else} \\
\>\>\>\textbf{if} $n=d$ \textbf{then} \\
\>\>\>\>\textbf{deliver} $L(s,d,F)$ \\
\>\>\>\textbf{elsif} $n$ if an entry point to $F'$ \textbf{then}\\
\>\>\>\>\textbf{add} $L(s,d,F')$ to $SQ$\\
\>\>\>\>\textbf{add} $R(s,d,F')$ to $SQ$ \\
\>\>\>\textbf{add} $L(s,d,F)$ to $SQ$ \\
\>\textbf{if} \textbf{receive} $R(s,d,F)$ \textbf{then} \\
\>\>/* handle similar to $L(s,d,F)$ */
\end{tabbing}
\caption{pseudocode for \emph{2FACE} at each node} \label{Fig2FACEcode}
\end{figure}

\begin{figure}
\centering \epsfig{figure=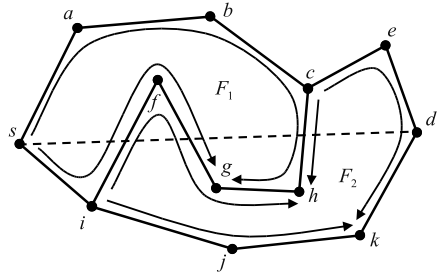,width=6cm,clip=}
\caption{Example of \emph{2FACE} operation. Nodes $s$, $i$ and $c$ are
entry points.} \label{fig2FACEexample}
\end{figure}

The operation of \emph{2FACE} is best understood with an
example. Consider the graph shown in Figure \ref{fig2FACEexample}. If
node $s$ has a message to transmit to node $d$, it sends $R(F_1)$ and
$L(F_1)$ tokens around face $F_1$. As a shorthand, we omit the source
and destination and just specify the face that the token traverses.
Nodes $a$ and $b$ forward $L(F_1)$ without other actions. Node $c$
also forwards $L$. However, $c$ is an entry point to an adjacent face
$F_2$. Thus, $c$ also sends $L(F_2)$ and $R(F_2)$.  Node $h$ forwards
$L(F_1)$ to $g$.  Meanwhile, $i$ also forwards $R(F_1)$ to $f$. Node
$i$ is another entry point to $F_2$ and it sends another pair of
tokens $L'(F_2)$ and $R'(F_2)$ to traverse $F_2$.  Node $f$ receives
$R(F_1)$ from $i$ and forwards it to $g$.  Node $g$ receives both
$R(F_1)$ and $L(F_1)$ and deletes them.  This completes the traversal
of $F_1$.

Notice that there are two pairs of tokens: $L$, $R$ and $L'$, $R'$
that traverse $F_2$. Token $L'$ is sent from $i$ to $h$ via $f$ and
$g$. At $h$ it meets $R$ where both tokens are destroyed. Token $L$ is
forwarded from $c$ to $e$ and then to $d$ where $L$ is
delivered. Token $L$ then continues to $k$ where it meets $R'$. Node
$k$ destroys the pair and completes the traversal of $F_2$.

\ \\ \textbf{Correctness proof.}

\begin{lemma}\label{JustOnce}
For each node $n$ bordering a face $F$ that intersects $(s,d)$ one of
the following happens exactly once: either (a) $n$ receives token
$T(s,d,F)$ where $T$ is either $R$ or $L$ and forwards it or (b) $n$
has a token, receives a token from the opposite direction and deletes
them both.
\end{lemma}

\Proof According to the algorithm, the token visits the node and
proceeds to the next one along the face, or the two tokens in the
opposing traversal directions meet at a node and disappear. Thus, to
prove the lemma, we have to show that each node bordering face $F$ is
reached and that it is visited only once. A sequence of adjacent nodes
of the face is a \emph{visited segment} if each node has been visited
at least once. A \emph{border} of a visited segment is a visited node
whose neighbor is not visited. By the design of the algorithm, a
border node always has a token to send to its neighbor that is not
visited. Because we assume reliable message transmission, eventually
the non-visited neighbor joins the visited segment. Thus, every node
in a face with a visited segment is eventually visited.

Observe that the face bordering $s$ has at least one visited segment:
the one that contains $s$ itself. Thus, every node in this face will
eventually be visited. Because graph $G$ is connected, there is a
sequence of adjacent faces intersecting $(s,d)$ from the face
bordering $s$ to the face bordering $d$. Adjacent faces share an entry
point. When an entry point is visited in one face, it sends a pair of
tokens around the adjacent face; that is, visiting an entry point
creates a visited segment in the adjacent edges. By induction, all
nodes in the sequence of adjacent faces are visited, including the
destination node.

Let us discuss if a token may penetrate a segment and arrive at an
interior (non-border) node. Observe that the computation of
\emph{2FACE} starts with a single segment consisting of the source
node. Thus, initially, there are no tokens inside any of the segments.
Assume there are no internal tokens in this computation up to some
step $x$. Let us consider the next step.  The token may penetrate the
segment only through a border node or through an interior entry
point. A token may arrive at a border node $b$ only from the border
node of another segment of the same face.  Because $b$ is a border
node, it already holds the token of the opposite traversal
direction. Thus, $b$ destroys both tokens and the received token does
not propagate to the interior nodes. Let us consider an entry point
node $e$. Because $e$ is interior to the segment, it was visited
earlier. Recall that a node is an entry point of two faces.  When an
entry point of a face receives a token, it creates a pair of tokens in
the other face. That is, once an entry point is visited, it becomes
visited in both faces. Since we assumed that there are no internal
tokens up to step $x$, $e$ cannot receive a token. By induction, a
token may not penetrate a segment. That is, each node bordering a face
is visited at most once.  This completes the proof of the lemma.  \QED

The below theorem follows from Lemma \ref{JustOnce}.

\begin{theorem}
Algorithm 2FACE guarantees the delivery of a message from $s$
to $d$.
\end{theorem}

According to Lemma \ref{JustOnce}, the total number of messages sent
in a computation is equal to the sum of the edges of the faces
intersecting $(s,d)$. The causally related messages propagate along a
path between $s$ and $d$. Hence, the following corollary.

\begin{corollary}
The worst case message complexity of 2FACE is $O(|E|)$ and time
complexity is $O(|V|)$.
\end{corollary}

\section{\emph{2FACE} Application and \\ Extensions}\label{secExt}

\textbf{Combining with greedy routing, using various traversals.}  For
efficiency a single direction face traversal may be combined with
greedy routing as in \emph{GFG} \cite{GFG}.  Algorithm \emph{2FACE}
can be used in a similar combination. We call the combined algorithm
\emph{G2FG}.  The message starts in greedy mode and switches to
\emph{2FACE} once it reaches a local minimum. Because multiple messages
traverse the graph simultaneously, unlike \emph{GFG}, once the message
switches to face traversal in \emph{G2FG} it continues in this mode
until the destination is reached. A technique similar to \emph{2FACE},
can be used for bi-directional void traversal \cite{Void}. The resultant
algorithm is \emph{2VOID}. \emph{2VOID} can also be combined with greedy routing
to form \emph{G2VG}.

Face traversal can be accelerated if each node stores its two-hop
neighbors as proposed by Datta et al \cite{Shortcut}. This method certainly
applies to \emph{2FACE}.

\ \\ \textbf{Following the shortest path.} The performance of
\emph{2FACE} can be further optimized if the source has multiple
messages to send to the destination; that is, there is a
\emph{session} between $s$ and $d$. The idea is to route the messages
in a single direction and only along the shorter route of the
face. This path is called \emph{preferred}. To enable this, the entry
point needs to be informed as to which traversal direction leads to
the shorter route to the destination. Algorithm \emph{2FACE} is
augmented by requiring each entry point to store the traversal
direction from which it was visited.  The first message in the session
is sent using augmented \emph{2FACE}.  When $d$ receives the message,
it sends a single \emph{traceback} message in the opposite
direction. This \emph{traceback} is to travel the preferred route in
reverse.  When an entry point gets a \emph{traceback}, it stores the
direction of its arrival and forwards it in the adjacent face in the
reverse direction from which it was first visited. Thus,
\emph{traceback} reaches the source and every entry point learns the
preferred direction to forward messages of the session henceforth.
Note that some of the entry points may lie on the preferred path
altogether.  A separate message sent using \emph{2FACE} has to inform
these entry points to discard the direction information. The last
message of the session that travels the preferred path has to make the
entry points forget about the preferred direction. Thus, following the
spirit of geometric routing, no information is stored at intermediate
nodes between sessions.

\begin{figure}[htb]
\centering\epsfig{figure=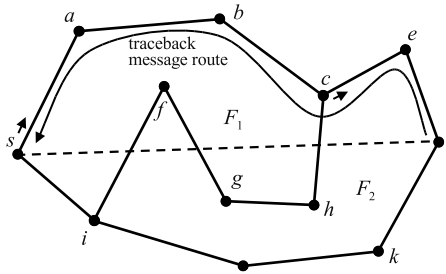,width=6cm,clip=}
\caption{The example traceback message route and traversal directions
stored at face entry points $s$ and $d$.}\label{figTraceback}
\end{figure}

Let us go back to the example in Figure \ref{fig2FACEexample} to
illustrate this idea.  For this example, the route of the traceback
message and the traversal directions for entry points $s$ and $d$ are
shown in Figure~\ref{figTraceback}. Also refer to the latter figure
for the subsequent discussion.  After the first \emph{2FACE} message,
the entry points $c$ and $i$ store the preferred token arrival
direction of $L$ and $R$ respectively. Destination $d$ receives the
left-hand-rule token first. Node $d$ sends the traceback token
$R$. When traceback $R$ reaches $c$, $c$ stores the preferred token
forwarding direction as $L$. Then, $c$ forwards the traceback token
$R$. When this token arrives at $s$, $s$ learns that the direction of
the preferred path is $L$. For the rest of the session, $s$ will send
messages to traverse $F_1$ using left-hand-rule until they reach
$c$. Node $c$ will forward them, also using left-hand-rule, until the
messages reach $d$. A message has to be sent using \emph{2FACE} to
inform entry point $i$ (which does not lie on the preferred path) that
entry point $i$ should not store direction information any longer.

\section{Performance Evaluation}\label{secSim}

\textbf{Simulation environment.} We programmed the algorithms described
in this paper using Java and Matlab.  We used sets of randomly
generated, connected unit-disk graphs starting with $40$ nodes and up
to $180$ nodes with $20$ node increments in an area of $2$ by $2$
units.

\begin{figure}[htb]
\begin{minipage}[b]{0.5\textwidth}
\centering \epsfig{figure=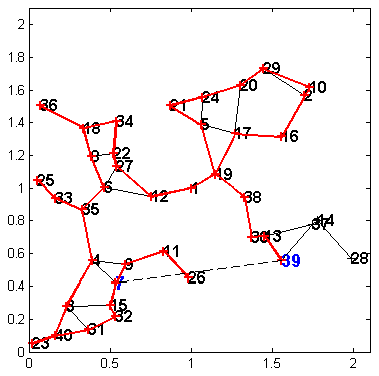,width=6cm,clip=}
\end{minipage}
\begin{minipage}[b]{0.5\textwidth}
\centering \epsfig{figure=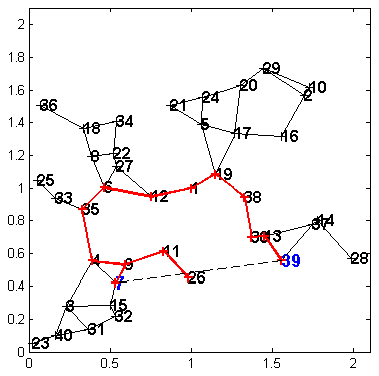,width=6cm,clip=}
\end{minipage}
\caption{Example routes selected by \emph{GFG} and \emph{G2FG} between
nodes $7$ and $39$ in a $40$-node graph} \label{figExampleRoutes}
\end{figure}

In each graph, the nodes were uniformly distributed over the area.  For
each set of generated graphs we used the connectivity unit $u$ ranging
from $0.9$ to $0.2$: a pair of nodes was connected by an edge if the
distance between them was less than $u$. Disconnected graphs were
discarded. For simulation we used an earlier version of \emph{FACE2}
where the token waits for its pair at an entry point. On the generated
graphs, the performance of this version and the one presented in the
paper are identical.  For each node density, $20$ graphs were generated
for the experiments.  Thus, the total number of graphs under
consideration was $160$ graphs.  For each graph, $20$ random node
pairs were selected as sources and destinations.  Example routes
selected by \emph{GFG} and \emph{G2FG} are shown in Figure
\ref{figExampleRoutes}. In this example, the preferred path distance
between the source and destination is $44$ hops for \emph{GFG} and
$14$ for \emph{G2FG}.

\begin{figure}[htb]
\begin{minipage}[b]{0.5\textwidth}
\centering \epsfig{figure=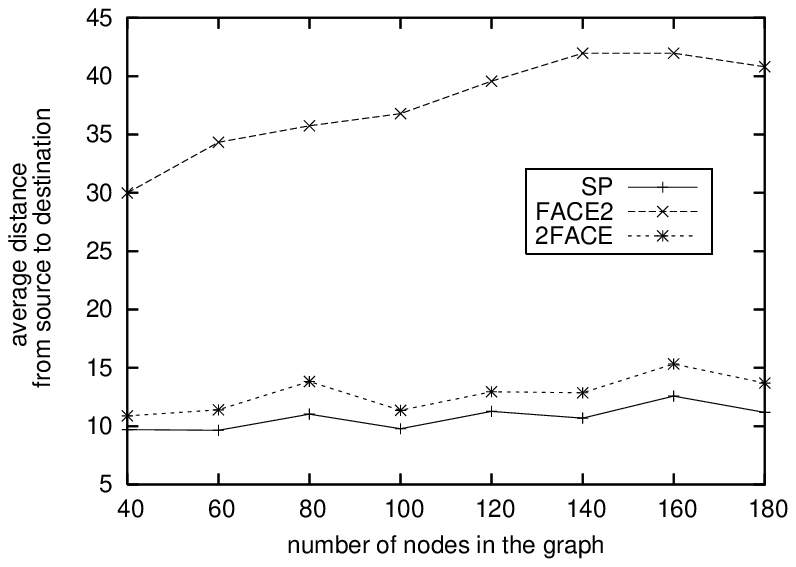,width=6cm,clip=}
\end{minipage}
\begin{minipage}[b]{0.5\textwidth}
\centering \epsfig{figure=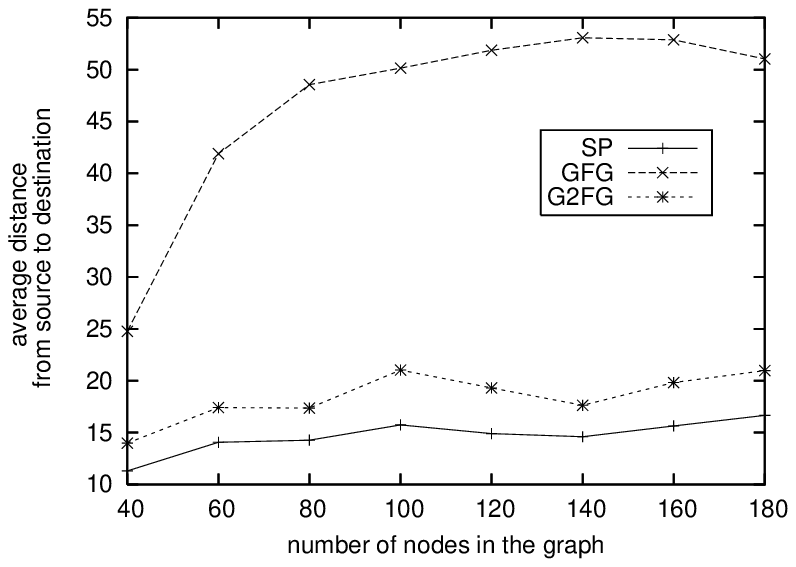,width=6cm,clip=}
\end{minipage}
\caption{Path length for single and bi-directional traversal 
in planar graphs.} \label{figPlanarPath}
\end{figure}

\begin{figure}[htb]
\begin{minipage}[b]{0.5\textwidth}
\centering \epsfig{figure=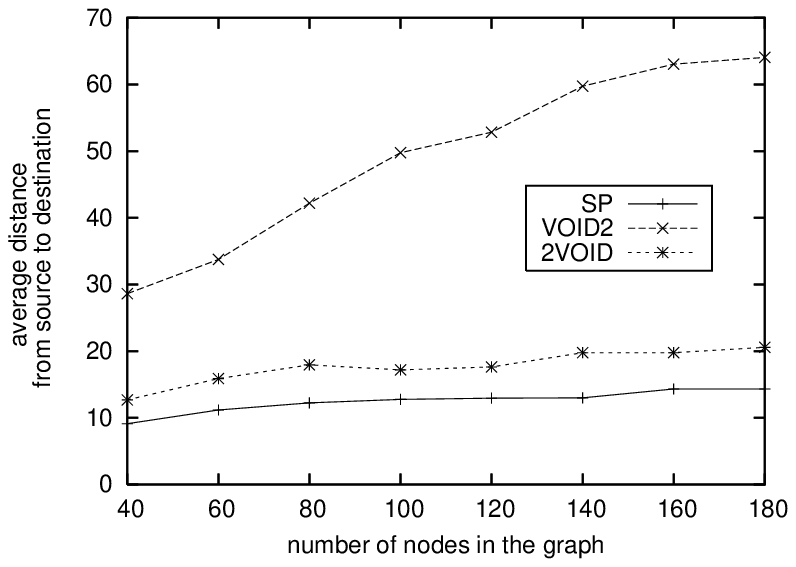,width=6cm,clip=}
\end{minipage}
\begin{minipage}[b]{0.5\textwidth}
\centering \epsfig{figure=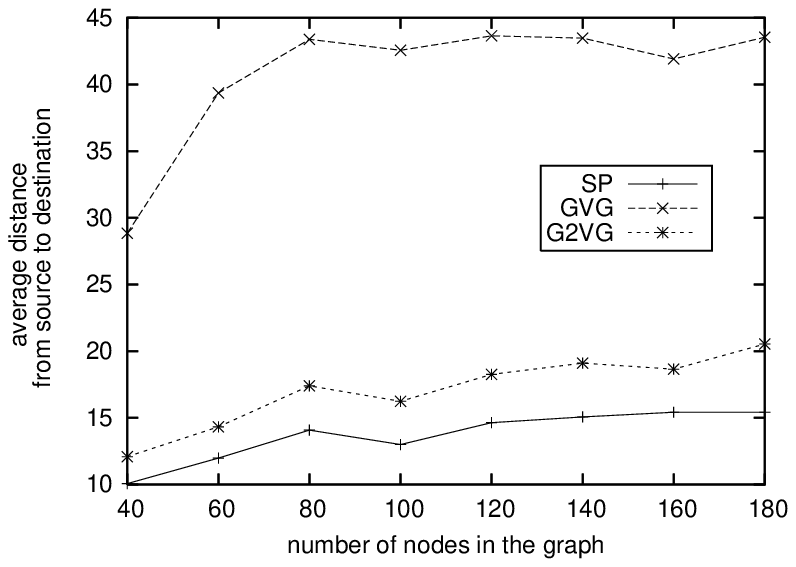,width=6cm,clip=}
\end{minipage}
\caption{Average path length for single and bi-directional
traversal in non-planar graphs.} \label{figVoidPath}
\end{figure}

\begin{figure}[htb]
\centering\epsfig{figure=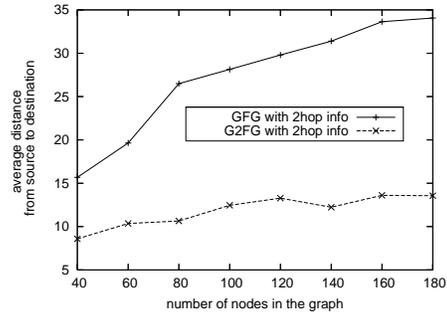,width=6cm,clip=}
\caption{Average path length for single and bi-directional traversal
with 2-hop shortcut optimization.} \label{fig2hop}
\end{figure}

\begin{figure}[htp]
\centering\epsfig{figure=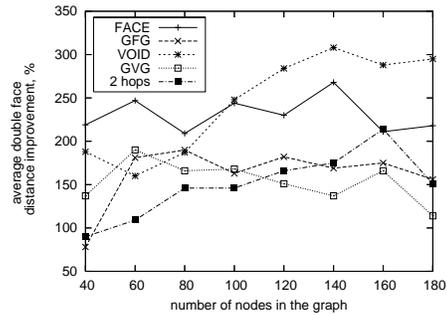,width=6cm,clip=}
\caption{Normalized improvement of bi-directional over
single-directional traversal.}\label{figPathDiff}
\end{figure}

\ \\ \textbf{Route length comparison.} We compare the optimal
(shortest) route with the route generated by the single-direction and
bi-direction traversal algorithm. For the bi-direction traversal we
used the preferred path for route length calculation.  For planar
graphs, we chose the more efficient \emph{FACE-2} for single-direction
traversal. For both kinds of traversals, we also compared the
performance of the greedy variants: \emph{GFG} and \emph{G2FG}.  The
results are shown in Figure \ref{figPlanarPath}.  We carried out the
same measurements for non-planar traversal algorithms. We chose
\emph{VOID-2} as a single-direction traversal algorithm. The results
are shown in Figure~\ref{figVoidPath}. We incorporated the 2-hop
shortcut optimization to traversal suggested by Datta et al
\cite{Shortcut}. The results are shown in Figure~\ref{fig2hop}. As the
results indicate, for all geometric routing algorithms studied, the
bi-directional traversal outperforms the single-directional one by a
factor of $2$ or $3$. Moreover, the path length of the bi-directional
traversal approaches the shortest (optimal) path in the graph.  To
highlight the performance improvement, we plotted the difference
between the paths selected by single- and bi-directional traversals
normalized to the bi-directional traversal path length. The plot is in
Figure~\ref{figPathDiff}. The plot does not show an observable trend,
but the path improvement is consistent across the graph densities and
and across different routing algorithms.

\begin{figure}[ht]
\begin{minipage}[b]{0.5\textwidth}
\centering \epsfig{figure=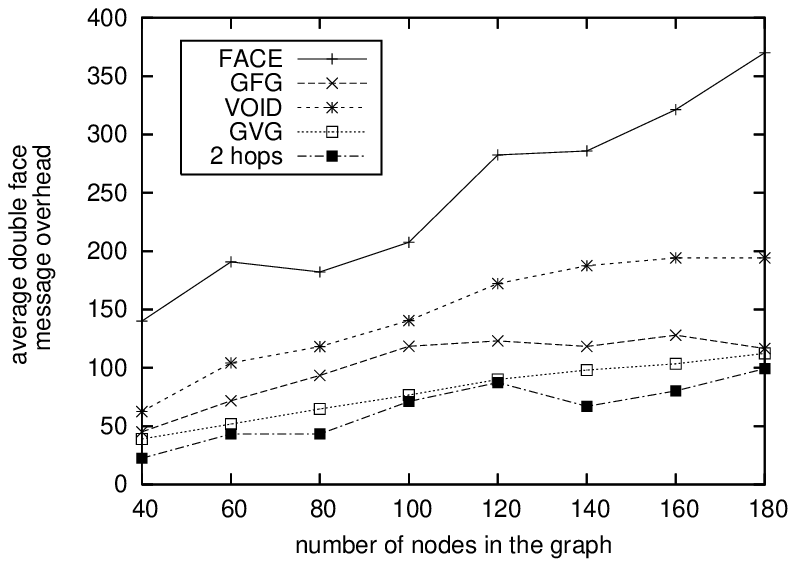,width=6cm,clip=}
\end{minipage}
\begin{minipage}[b]{0.5\textwidth}
\centering \epsfig{figure=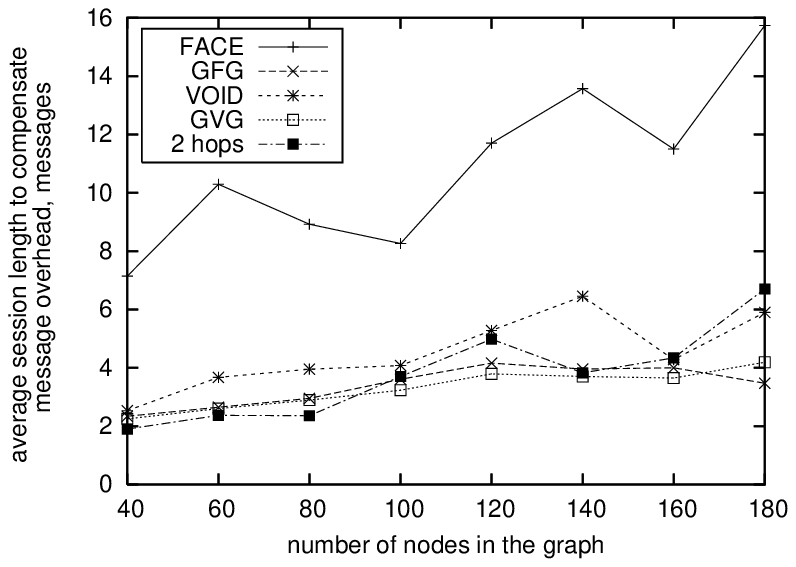,width=6cm,clip=}
\end{minipage}
\caption{Message overhead and number of messages
to compensate the overhead} \label{figOverhead}
\end{figure}

The concurrent message transmission in bi-directional traversal
requires more messages than in in single direction. We quantify this
message overhead in Figure~\ref{figOverhead}. In the first graph, we
plat the message difference between the two traversal types. Observe
that the preferred route selected by bi-directional traversal is
several times shorter on average. Thus, if source have multiple
messages to send to the destination, the subsequent messages can use
this shorter route as opposed to the route selected by the
single-directional traversal. The second graph in
Figure~\ref{figOverhead} indicates how many messages per session it
takes to compensate this overhead. The figure indicates that for some
routing algorithms the overhead may be recouped by a session as short as
just two messages.

\section{Conclusion} \label{secEnd}
In this paper, we proposed the bi-directional face traversal to
improve geometric routing efficiency. The results show that the
proposed algorithm provides significant performance improvement over
existing single-face traversal.  Moreover, the bi-directional
traversal addresses one of the major drawbacks of geometric routing:
its inconsistency due to selection of disadvantageous routes. The
proposed technique is simple to implement. The authors are hopeful
that it will find its way into the practical implementation of routing
algorithms.

\bibliographystyle{plain}
\bibliography{2face}
\label{docend}
\end{document}